\documentclass[conference]{IEEEtran}
\IEEEoverridecommandlockouts

\usepackage{titlesec}
\let\subparagraph\relax
\titlespacing{\section}{0pt}{6pt plus 1pt minus 1pt}{4pt plus 1pt minus 1pt} 
\usepackage{cite}
\usepackage{amsmath,amssymb,amsfonts}
\usepackage{algpseudocode}
\usepackage{float}
\usepackage[ruled,vlined]{algorithm2e}
\usepackage{graphicx}
\usepackage{textcomp}
\usepackage{xcolor}
\def\BibTeX{{\rm B\kern-.05em{\sc i\kern-.025em b}\kern-.08em
    T\kern-.1667em\lower.7ex\hbox{E}\kern-.125emX}}
\begin{document}

\title{
Beamforming Design Via GNN in mmWave Cell-Free Massive MIMO Using Sub-6\,GHz CSI  
}

\author{%
\IEEEauthorblockN{%
Sina Tavakolian$^{\star}$,
Abolfazl Zakeri$^{\star}$,
Ahmed Alkhateeb$^{\dagger}$,
Markku Juntti$^{\star}$, 
and Nhan Thanh Nguyen$^{\star}$,
}
\IEEEauthorblockA{%
$^{\star}$ Centre for Wireless Communications, University of Oulu, P.O.Box 4500, FI-90014, Finland \\
  $^{\dagger}$School of Electrical, Computer, and Energy Engineering, Arizona State University, AZ, USA\\
\{sina.tavakolian, abolfazl.zakeri, nhan.nguyen, markku.juntti\}@oulu.fi, alkhateeb@asu.edu
}
}

\maketitle

\begin{abstract}
Beamforming methods in millimeter-wave (mmWave) cell-free massive multiple-input multiple-output (CFmMIMO) systems require accurate channel state information (CSI), whose acquisition entails significant training overhead. 
This paper shows that fully digital cell-free mmWave beamforming can be effectively learned from sub-6\,GHz CSI
using a graph neural network (GNN). Specifically, we represent a CFmMIMO system as a wireless graph, and The GNN is trained to approximate beamformers that maximize the downlink sum-rate, based on the available sub-6\,GHz CSI. A message passing mechanism is proposed to capture inter-user interference and inter-basestation cooperation across different network topologies. Simulation results demonstrate that the proposed sub-6 GHz–assisted GNN-based beamformer achieves competitive and often superior sum-rate performance compared to classical baselines that rely on full mmWave CSI.
\end{abstract}


\section{Introduction}
Cell-free massive multiple-input multiple-output (CFmMIMO) has emerged as a promising architecture for future wireless networks, where multiple distributed base stations (BSs) jointly serve user equipments (UEs) and are all connected to a central processing unit (CPU)~\cite{cellfree_vs_smallcells, scalable_cellfree}.
When operating at millimeter-wave (mmWave) frequencies, however, beamforming becomes essential due to the severe propagation loss and the use of large antenna arrays~\cite{heath2016}. 
Conventional mmWave beamforming designs typically rely on channel state information (CSI). 
However, acquiring mmWave CSI in CFmMIMO systems entails significant training overhead.
\\\indent
Several works have explored the use of out-of-band (OOB) side information, particularly sub-6\,GHz CSI~\cite{OOB_latency,alrabeiah2020sub6map,tavakolian2026kd_mmwave,Deng_JSAC_2025,Sub-6GHz_Assisted_mmWave_Hybrid_Beamforming_with_Heterogeneous_Graph_Neural_Network}. 
Such sub-6\,GHz channels can be estimated much more efficiently than mmWave channels using conventional pilot-based procedures, making them a useful source of side information for mmWave beamforming design.
Early studies exploited cross-band spatial correlation for beam selection~\cite{OOB_latency}, while more recent machine learning (ML)-based approaches use neural networks to predict mmWave beams directly from sub-6\,GHz CSI~\cite{alrabeiah2020sub6map,tavakolian2026kd_mmwave}. 
Particularly, graph neural network (GNN)-based methods have also been explored for OOB-assisted mmWave beamforming~\cite{Deng_JSAC_2025,Sub-6GHz_Assisted_mmWave_Hybrid_Beamforming_with_Heterogeneous_Graph_Neural_Network}. 
However, existing OOB-assisted approaches, including~\cite{Deng_JSAC_2025,Sub-6GHz_Assisted_mmWave_Hybrid_Beamforming_with_Heterogeneous_Graph_Neural_Network}, primarily focus on single-BS or multi-cell systems. Specifically,~\cite{Deng_JSAC_2025} considers only a single BS, whereas in ~\cite{Sub-6GHz_Assisted_mmWave_Hybrid_Beamforming_with_Heterogeneous_Graph_Neural_Network}, each UE is assigned to one predetermined mmWave BS, allowing the graph edges to be a priori classified as desired or interfering links; moreover, they assume partial mmWave CSI is also required. These assumptions do not apply to CFmMIMO systems, where multiple distributed BSs jointly serve each UE without a fixed desired/interfering-link partition. Therefore, beamforming design for CFmMIMO systems using \textit{solely} sub-6\,GHz CSI remains unexplored.
\\\indent 
To address the above gap, in this work, we study downlink beamforming design in a CFmMIMO system where only uplink sub-6\,GHz CSI is available at the CPU. We formulate the beamforming problem as an ML task and develop a GNN framework that learns digital beamformers directly from sub-6\,GHz CSI.
The CFmMIMO system is modeled as a graph, and
the proposed GNN performs customized message passing over this graph to capture both multi-user interference at each BS and cooperative transmission across BSs.
Importantly, the proposed framework is trained end-to-end to directly maximize the sum-rate, without relying on supervised labels (e.g., optimal beamforming vectors).
Moreover, the graph-based formulation allows the learned model to naturally adapt to different and \textit{dynamic} network topologies, e.g., varying number of users.
Numerical results demonstrate that the learned beamformers achieve performance close to the baselines that rely on full mmWave CSI, including minimum mean-square error (MMSE) beamforming. Moreover, the proposed framework maintains good performance across varying network topologies and remains robust when only partial sub-6\,GHz CSI is available.







\section{System Model and Problem Formulation}
\label{sec:system_model}
\subsection{System Model}
We consider a downlink cell-free dual-band single-carrier communications system. The system consists of $B$ distributed BSs connected to a CPU and jointly serving $U$ UEs. Each BS is equipped with two co-located antenna arrays: a sub-6\,GHz array with $N^{\mathrm{s}}$ antennas, and a mmWave array with $N^{\mathrm{mW}}$ antennas. Each UE employs a single antenna at both sub-6\,GHz and mmWave frequency bands.
Downlink data transmission is carried out in the mmWave band. 
%

%
%
We adopt the geometric channel model in~\cite{alrabeiah2020sub6map} for both the sub-6\,GHz uplink and the mmWave downlink. Although the propagation environment is wideband and includes multiple paths with distinct delays, as described in~\cite[Eq.~(3)]{alrabeiah2020sub6map}, the following processing and analysis are performed over a single subcarrier for simplicity. Let ${\mathbf{h}_{b,u}^{\mathrm{s}} \in \mathbb{C}^{N^{\mathrm{s}}\times1}}$ and ${\mathbf{h}_{b,u}^{\mathrm{mW}} \in \mathbb{C}^{N^{\mathrm{mW}}\times1}}$ denote the corresponding sub-6\,GHz and mmWave channel vectors, respectively. Their parameters are generated through 3D ray-tracing simulations, which account for realistic frequency-dependent propagation effects and preserve spatial consistency across the two frequency bands, as detailed in~\cite{alrabeiah2020sub6map}.

Let $\mathbf{s}_b = [s_{b,1}, s_{b,2}, \dots, s_{b,U}]^\mathsf{T} \in \mathbb{C}^{U \times 1}$ denote the data symbol vector transmitted by BS~$b$, where $s_{b,u}$ is the data symbol intended for UE~$u \in \mathcal{U}$ and $\mathbb{E}[\mathbf{s}_b \mathbf{s}_b^\mathsf{H}] = \mathbf{I}_U$. 
Each BS employs a beamformer $\mathbf{F}_b \in \mathbb{C}^{N^{\mathrm{mW}}\times U}$ to precode the transmitted signal, resulting in the transmit vector $\mathbf{F}_b \mathbf{s}_b$. 
The received signal at UE~$u \in \mathcal{U}$ is then given by
\begin{equation}
y_u = \sum_{b=1}^{B} {\mathbf{h}_{b,u}^{\mathrm{mW}}}^{\mathsf{H}} \mathbf{F}_b \mathbf{s}_b + n_u,
\label{eq:rx_signal}
\end{equation}
where $n_u \sim \mathcal{CN}(0,\sigma^2)$ denotes the additive white Gaussian noise.
Assuming that symbol $s_{b,u}$ is intended for UE $u$ from BS $b$, we can write:
\begin{equation}
y_u =
\underbrace{\sum_{b=1}^{B} {\mathbf{h}_{b,u}^{\mathrm{mW}}}^{\mathsf{H}} 
\mathbf{f}_{b,u} s_{b,u}}_{\text{Desired signal}}
+
\underbrace{\sum_{b=1}^{B} \sum_{\substack{j \in \mathcal{U} \\ j \ne u}}
{\mathbf{h}_{b,u}^{\mathrm{mW}}}^{\mathsf{H}} 
\mathbf{f}_{b,j} s_{b,j}}_{\text{Inter-UE interference}}
+ n_u ,
\label{eq:rx_signal_decomposed}
\end{equation}
where $\mathbf{f}_{b,u}$ denotes the $u$-th column of $\mathbf{F}_b$.
The downlink SINR at UE~$u$ is therefore given as
\begin{align}
\gamma_u =
\frac{\left|\sum_{b=1}^{B} {\mathbf{h}_{b,u}^{\mathrm{mW}}}^{\mathsf{H}} \mathbf{f}_{b,u}\right|^2}
{\underset{j \in \mathcal{U} , j \ne u}{\sum}
\left|\sum_{b=1}^{B} {\mathbf{h}_{b,u}^{\mathrm{mW}}}^{\mathsf{H}} \mathbf{f}_{b,j}\right|^2 + \sigma^2}.
\label{eq:sinr}
\end{align}
Accordingly, the achievable downlink rate of UE~$u$ is ${R_u = \log_2(1+\gamma_u)}$.
The sum-rate of the system is then given~by
\begin{equation}
R^{\mathrm{sum}} = \sum_{u=1}^{U} R_u.
\label{eq:sumrate}
\end{equation}

\subsection{Problem Formulation}
The goal is to design the fully digital beamformers $\{\mathbf{F}_b\}_{b=1}^{B}$ that maximize the downlink sum-rate in~\eqref{eq:sumrate}. The beamforming design problem can be written as
\begin{subequations}
\label{eq:main_problem}
\begin{align}
\underset{\{\mathbf{F}_b\}_{b=1}^{B}}{\mbox{maximize}}
\quad & R^{\mathrm{sum}} \label{eq:main_problem_obj}\\
\mbox{subject to}
\quad 
& \|\mathbf{F}_b\|_F^2 = P_b, \quad \forall\, b,
\label{eq:power_constraint_eq}
\end{align}
\end{subequations}
where $P_b$ denotes the power budget at BS $b$. The transmit power constraint is enforced as an equality in~\eqref{eq:power_constraint_eq} because, for sum-rate maximization, the optimal solution often utilizes the fully available transmit power~\cite{bjornson2017massive}.
\\\indent 
Solving problem~\eqref{eq:main_problem} directly would require the mmWave downlink CSI, which is supposed to be \textit{unavailable} in our considered scenario. Instead, only the uplink sub-6\,GHz CSI is available at the CPU, and we aim to solve the design problem centrally using this information.  This assumption is motivated by the fact that the fully digital sub-6\,GHz arrays provide direct baseband access to all antenna signals, enabling efficient pilot-based CSI estimation with low overhead, whereas estimating the mmWave channel typically involves extensive beam training~\cite{song2021joint}.
\\\indent
The uplink sub-6\,GHz CSI is collected as
\begin{equation}
\mathbf{H}^{\mathrm{s}}
=
\big[\mathbf{H}_1^{\mathrm{s}},\dots,\mathbf{H}_B^{\mathrm{s}}\big]
\in \mathbb{C}^{B\times U\times N^{\mathrm{s}}},
\label{eq:global_sub6}
\end{equation}
where
\begin{equation}
\mathbf{H}_b^{\mathrm{s}}
=
\begin{bmatrix}
(\mathbf{h}_{b,1}^{\mathrm{s}})^\top\\
\vdots\\
(\mathbf{h}_{b,U}^{\mathrm{s}})^\top
\end{bmatrix}
\in \mathbb{C}^{U\times N^{\mathrm{s}}}.
\label{eq:local_sub6}
\end{equation}
To perform beamforming design in~\eqref{eq:main_problem} without the mmWave CSI, we aim to learn a mapping from the available sub-6\,GHz CSI to the digital beamformers, i.e.,
\begin{equation}
\Phi^{\mathrm{s}}\!\left(\mathbf{H}^{\mathrm{s}}\right)
=
\{\mathbf{F}_b\}_{b=1}^{B},
\label{eq:mapping}
\end{equation}
where $\Phi^{\mathrm{s}}(\cdot)$ is learned such that the resulting beamformer maximizes the sum-rate in~\eqref{eq:sumrate} while satisfying the per-BS power constraint in~\eqref{eq:power_constraint_eq}.
Finding the mapping $\Phi^{\mathrm{s}}(\cdot)$ in closed form is analytically intractable~\cite{alrabeiah2020sub6map}. This motivates the use of a data-driven ML framework that approximates $\Phi^{\mathrm{s}}(\cdot)$ directly from sub-6\,GHz channels. GNNs provide a suitable ML architecture for this problem, as they can effectively model interactions among multiple BSs and UEs whose wireless channels jointly determine the beamforming decisions. To this end, we develop a GNN-based framework that learns the mapping in~\eqref{eq:mapping} and generalizes to different network topologies, as described next.
\section{Proposed GNN-Based Framework}
\label{subsec:graph_construction}
We model the considered CF-mMIMO system as a \emph{wireless graph}, defined as $\mathcal{G} := (\mathcal{V}, \mathcal{E}, \mathcal{T}_{\mathcal{V}})$, 
where the node set is given by $\mathcal{V} = \mathcal{V}_{\mathrm{BS}} \cup \mathcal{V}_{\mathrm{UE}}$, consisting of the BS nodes $\mathcal{V}_{\mathrm{BS}} = \{1,\dots,B\}$ and UE nodes $\mathcal{V}_{\mathrm{UE}} = \{1,\dots,U\}$. The edge set $\mathcal{E} = \{(b,u) \mid b \in \mathcal{V}_{\mathrm{BS}},\; u \in \mathcal{V}_{\mathrm{UE}} \}$
captures all communication links between BSs and UEs. The node-type mapping is specified by $\mathcal{T}_{\mathcal{V}} = \{\mathrm{BS}, \mathrm{UE}\}$. For each edge $(b,u) \in \mathcal{E}$, we assign a feature vector extracted from the corresponding sub-6\,GHz channel $\mathbf{h}_{b,u}^{\mathrm{s}}$. Furthermore, each BS and UE node is associated with a learnable embedding vector, denoted by $\boldsymbol{\rho}_b \in \mathbb{R}^{d}$ and $\boldsymbol{\rho}_u \in \mathbb{R}^{d}$, respectively. These embeddings are later optimized jointly with the GNN parameters. The parameter $d$ represents the dimensionality of all feature and embedding vectors.
\subsection{GNN Architecture}
\label{subsec:gnn_architecture}
With the graph representation described above, we aim to design a GNN that generates the beamformers efficiently. 
GNNs operate by iteratively propagating information across the graph through message passing, enabling each element, i.e., each node or edge, to incorporate information from its neighbors~\cite{gnngeneral,gilmer2017neural}.
Following this principle, the proposed GNN method consists of three phases: 

1) \textit{Initialization (Phase 1):}
Let $\mathbf{z}_b^{(t)}$ and $\mathbf{z}_u^{(t)}$ denote the hidden states of BS~$b$ and UE~$u$ at message-passing iteration~$t$, respectively. 
Similarly, let $\mathbf{e}_{b,u}^{(t)}$ represent the hidden state of edge $(b,u)$.
At the initialization step, i.e., $t=0$, the node hidden states are set to the learnable embeddings introduced in the previous subsection, while the edge hidden states are obtained from the sub-6\,GHz channel features, i.e., 
\begin{equation}
\label{eq:initializationeq}
\mathbf{z}_b^{(0)} = \boldsymbol{\rho}_b, \quad
\mathbf{z}_u^{(0)} = \boldsymbol{\rho}_u, \quad
\mathbf{e}_{b,u}^{(0)} = \varphi_{\mathrm{edge}}^{\mathrm{init}}\!\left(\mathbf{h}_{b,u}^{\mathrm{s}}\right),
\end{equation}
where $\varphi_{\mathrm{edge}}^{\mathrm{init}}(\cdot)$ is a learnable affine mapping, i.e., a fully linear network without an activation layer, that projects the sub-6\,GHz channel features into a $d$-dimensional embedding space. 
The initialized hidden states $\mathbf{z}_b^{(0)}$, $\mathbf{z}_u^{(0)}$, and $\mathbf{e}_{b,u}^{(0)}$ serve as the starting representations that will be progressively refined through the message passing process described next.

2) \textit{Message Passing (Phase 2):}
At each iteration $t$, the hidden states of the edges are updated by aggregating information from neighboring BS-UE edges. 
Our message passing mechanism is inspired by the graph attention layer with edge features (EdgeGATConv) in~\cite{scene}. 
However, unlike~\cite{scene}, which aggregates edge information into node embeddings, we adapt the attention mechanism so that each edge directly aggregates information from its neighboring edges. For an edge $(b,u)$, we consider two neighborhoods: 
(i) the edges connected to the same BS~$b$, and 
(ii) the edges connected to the same UE~$u$. 
We denote the BS-side neighborhood by $\mathcal{E}(b)=\{(b,u')\,|\,u'\in\mathcal{N}(b)\}$ and the UE-side neighborhood by $\mathcal{E}(u)=\{(b',u)\,|\,b'\in\mathcal{N}(u)\}$. In the following equations, we use the superscripts ``$\mathrm{BS}$" and ``$\mathrm{UE}$" for parameters associated with the BS and UE, respectively.

We first compute attention coefficients that quantify the relevance of neighboring edges. 
For the BS-side neighborhood, we define the attention weights as
\begin{align}
\alpha_{b,u \leftarrow v}^{(t)}
&=
\mathrm{softmax}\hspace{1mm}\Big(
\mathbf{a}_{\mathrm{BS}}^{\top}
\hspace{1mm}\sigma\big(
\mathrm{concat}\hspace{0.5mm}\big(
\mathbf{W}^{\mathrm{BS}}_e \mathbf{e}_{b,u}^{(t)},\,
\mathbf{W}^{\mathrm{BS}}_e \mathbf{e}_{b,v}^{(t)}, \nonumber\\
&\hspace{3.5cm}
\mathbf{W}^{\mathrm{BS}}_b \mathbf{z}_b^{(t)}
\big)
\big)
\Big),
\label{eq:bs_attention}
\end{align}
where $\mathbf{a}_{\mathrm{BS}}$ is an attention vector used to score the importance of neighboring edges in the BS-side aggregation.
Here, $\mathbf{W}^{\mathrm{BS}}_e,\mathbf{W}^{\mathrm{BS}}_b\in\mathbb{R}^{d\times d}$ are trainable projection matrices that transform the edge and BS-node features into a common latent space before attention is computed.
The operator $\mathrm{concat}(\cdot)$ denotes concatenation, and $\sigma(\cdot)$ denotes the leaky rectified linear unit (LeakyReLU) activation function.
Using these attention weights, we compute the BS-side aggregated message as
\begin{align}
\mathbf{m}^{\mathrm{BS}}_{b,u}
&=
\!\!\!\!\!\sum_{(b,v)\in\mathcal{E}(b)}
\!\!\!\alpha_{b,u\leftarrow v}^{(t)}
\Big(
\mathbf{W}^{\mathrm{BS}}_e\mathbf{e}_{b,v}^{(t)}
+
\mathbf{W}^{\mathrm{BS}}_b\mathbf{z}_b^{(t)}
+
\mathbf{W}^{\mathrm{BS}}_u\mathbf{z}_u^{(t)}
\Big),
\label{eq:bs_message}
\end{align}
where $\mathbf{W}^{\mathrm{BS}}_u\in\mathbb{R}^{d\times d}$ transforms the UE-node features into the BS-side latent space before aggregation.

Analogously, for the UE-side neighborhood we compute UE-side attention coefficients for edge $(b,u)$ as
\begin{align}
\beta_{b \leftarrow w,u}^{(t)}
&=
\mathrm{softmax}\hspace{1mm}\Big(
\mathbf{a}_{\mathrm{UE}}^{\top}
\hspace{1mm}\sigma\big(
\mathrm{concat}\hspace{0.5mm}\big(
\mathbf{W}^{\mathrm{UE}}_e \mathbf{e}_{b,u}^{(t)},\,
\mathbf{W}^{\mathrm{UE}}_e \mathbf{e}_{w,u}^{(t)}, \nonumber\\
&\hspace{3.6cm}
\mathbf{W}^{\mathrm{UE}}_u \mathbf{z}_u^{(t)}
\big)
\big)
\Big),
\label{eq:ue_attention}
\end{align}
and we compute the corresponding message as
\begin{align}
\mathbf{m}^{\mathrm{UE}}_{b,u}
&=
\!\!\!\!\!\sum_{(w,u)\in\mathcal{E}(u)}
\!\!\!\!\!\beta_{b \leftarrow w,u}^{(t)}
\Big(
\mathbf{W}^{\mathrm{UE}}_e\mathbf{e}_{w,u}^{(t)}
+
\mathbf{W}^{\mathrm{UE}}_b\mathbf{z}_b^{(t)}
+
\mathbf{W}^{\mathrm{UE}}_u\mathbf{z}_u^{(t)}
\Big),
\label{eq:ue_message}
\end{align}
where $\mathbf{W}^{\mathrm{UE}}_b,\mathbf{W}^{\mathrm{UE}}_e,\mathbf{W}^{\mathrm{UE}}_u\in\mathbb{R}^{d\times d}$ are trainable matrices.
We then fuse the two aggregated messages in~\eqref{eq:bs_message} and~\eqref{eq:ue_message} as
\begin{equation}
\mathbf{m}_{b,u}^{(t)}
=
\frac{1}{2}
\Big(
\mathbf{m}^{\mathrm{BS}}_{b,u}
+
\mathbf{m}^{\mathrm{UE}}_{b,u}
\Big).
\label{eq:edge_reduce}
\end{equation}
Finally, we update the hidden state of edge $(b,u)$ using a residual rule such that
\begin{equation}
\mathbf{e}_{b,u}^{(t+1)}
=
\mathbf{W}^{\mathrm{upd}}\mathbf{e}_{b,u}^{(t)}
+
\mathbf{m}_{b,u}^{(t)}.
\label{eq:edge_update}
\end{equation}
Here $\mathbf{W}^{\mathrm{upd}}$ is a learnable matrix that linearly transforms the current edge state, forming the residual component of the update.
Note that in equations \eqref{eq:bs_attention}--\eqref{eq:edge_update}, we omit the bias terms, following the design of EdgeGATConv in~\cite{scene}.
We stack $T$ such message passing layers to progressively refine the edge representations.

3) \textit{Readout (Phase 3):}
After $T$ message passing layers, for a given edge $(b,u)$, we collect information from two groups of neighboring edges and summarize them through mean aggregation. Specifically, one group include the edges connected to the same BS $b$, namely $(b,u')\in\mathcal{E}(b)$ with $u'\neq u$, and the other group is for the edges connected to the same UE $u$, namely $(b',u)\in\mathcal{E}(u)$ with $b'\neq b$.
We construct the contextual feature vector as
\begin{equation}
\mathbf{c}_{b,u}
=
\mathrm{concat}\!\left(
\mathbf{e}^{(T)}_{b,u},
\mathbf{c}^{\mathrm{intra}}_{b,u},
\mathbf{c}^{\mathrm{inter}}_{b,u}
\right),
\end{equation}
where $\mathbf{c}^{\mathrm{intra}}_{b,u}$ and $\mathbf{c}^{\mathrm{inter}}_{b,u}$ are the \emph{mean} aggregated edge features from the BS-side and UE-side neighborhoods, obtained as
\begin{equation}
\mathbf{c}^{\mathrm{intra}}_{b,u}
=
\frac{1}{|\mathcal{E}(b)|-1}
\sum_{\substack{(b,u')\in\mathcal{E}(b)\\ u'\neq u}}
\mathbf{e}^{(T)}_{b,u'},
\end{equation}
\begin{equation}
\mathbf{c}^{\mathrm{inter}}_{b,u}
=
\frac{1}{|\mathcal{E}(u)|-1}
\sum_{\substack{(b',u)\in\mathcal{E}(u)\\ b'\neq b}}
\mathbf{e}^{(T)}_{b',u}.
\end{equation}
Thus, $\mathbf{c}_{b,u}$ captures the direct BS-UE relation together with mean-aggregated contextual information from neighboring BS and UE connections.

We then transform these contextual features into beamforming vectors $\{\mathbf{f}_{b,u}\}$ using a mapping $\varphi^{\mathrm{out}}(\cdot)$ implemented by
a multilayer perceptron (MLP), i.e.,
\begin{equation}
\mathbf{f}_{b,u}=\varphi^{\mathrm{out}}\!\left(\mathbf{c}_{b,u}\right)
\in\mathbb{C}^{N^{\mathrm{mW}}}.
\end{equation}
We form the digital beamformer at BS $b$ as ${ \mathbf{F}_b = \big[\mathbf{f}_{b,1},\dots,\mathbf{f}_{b,U}\big] }$.
Finally, we normalize $\mathbf{F}_b$ to satisfy the per-BS transmit power
constraint in~\eqref{eq:power_constraint_eq} as
\begin{equation}
\label{eq:normdig}
\mathbf{F}_b
\;\leftarrow\;
\sqrt{\frac{P_b}{\|\mathbf{F}_b\|_F^2}}
\,\mathbf{F}_b .
\end{equation}

\subsection{Offline Training and Online Operation}
\label{subsec:training}

\textit{Offline Training:}
The proposed GNN is trained in a self-supervised manner by directly optimizing the communication performance metric, without relying on labelled beamforming targets. We train the network to maximize the downlink sum-rate defined in~\eqref{eq:sumrate}, and the loss function is defined as the negative sum-rate, i.e., $\mathcal{L}=- R^{\mathrm{sum}}$. 
The overall training procedure of the proposed framework is summarized in Algorithm~\ref{alg:training}.
After constructing the dataset from the sub-6\,GHz CSI as shown in steps 1--5, the GNN performs a forward pass following the architecture in Section~\ref{subsec:gnn_architecture}, where node and edge states are initialized as in step 9 and refined through $T$ message-passing layers, following steps 10--14. 
The readout stage generates the beamforming vectors, which are normalized to satisfy the per-BS power constraints, according to steps 15--18. 
Given the beamformers obtained from the forward pass, the achievable rates and the loss function are computed.
For this purpose, the mmWave CSI is required \textit{only during the training phase}.
Let the global mmWave CSI be
\begin{equation}
\mathbf{H}^{\mathrm{mW}}=
\big[\mathbf{H}_1^{\mathrm{mW}},\dots,\mathbf{H}_B^{\mathrm{mW}}\big]
\in \mathbb{C}^{B\times U\times N^{\mathrm{mW}}},
\end{equation}
where each block $\mathbf{H}_b^{\mathrm{mW}}\in\mathbb{C}^{U\times N^{\mathrm{mW}}}$ contains the mmWave channels from BS~$b$ to all UEs.
Using $\mathbf{H}^{\mathrm{mW}}$ together with the predicted beamformers $\{\mathbf{F}_b\}_{b=1}^{B}$, the loss is evaluated, following Steps 19--20. 
The model parameters are then updated iteratively over minibatches and training epochs to minimize the loss, as shown in Steps 21--22.





\textit{Online Operation:}
At inference time (i.e., during model deployment), \textit{only} the sub-6\,GHz CSI $\mathbf{H}^{\mathrm{s}}$ is required to feed the (pretrained) model. Given a new graph constructed from $\mathbf{H}^{\mathrm{s}}$, the trained GNN generates the beamforming vectors $\{\mathbf{F}_b\}_{b=1}^{B}$ through a single forward pass.

The complexity of the proposed method is dominated by the GNN forward pass, particularly the message passing and aggregation over BS--UE edges, which scales with the number of links and their neighborhoods.
However, the main complexity reduction is at the system level, where the proposed approach avoids explicit mmWave CSI acquisition and instead infers beamformers from sub-6\,GHz CSI, which requires significantly lower training and feedback overhead.

\SetKwInput{KwData}{Parameters}

\begin{algorithm}[!t]
\small
\caption{Training of the Proposed Framework}
\label{alg:training}
\LinesNumbered
\KwIn{Global sub-6\,GHz CSI $\mathbf{H}^{\mathrm{s}}$; global mmWave CSI $\mathbf{H}^{\mathrm{mW}}$; 
per-BS powers $\{P_b\}_{b=1}^{B}$. 
}
\KwOut{Trained GNN parameters $\boldsymbol{\Theta}$ and predicted digital beamformers $\{\mathbf{F}_b\}_{b=1}^{B}$.}

\KwData{$B$; $U$; $T$; total number of graph samples $N$; batch size $M$; learning rate $\eta$; optimizer $\mathtt{Opt}$; number of training epochs $E$.}



\tcp{Dataset construction}
$\mathcal{D}\gets\emptyset$\;
\For{$i\gets1$ \KwTo $N$}{
Construct $\mathcal{G}^{(i)}=(\mathcal{V}^{(i)},\mathcal{E}^{(i)},\mathcal{T}_{\mathcal{V}})$ from the sub-6\,GHz CSI $\mathbf{H}^{\mathrm{s}}$ according to Section~\ref{subsec:graph_construction}\;
Store $\big(\mathcal{G}^{(i)}\big)$ in $\mathcal{D}$\;
}

Split $\mathcal{D}$ into training/test sets and form batches of size $M$\;
Initialize GNN parameters $\boldsymbol{\Theta}$;

\tcp{End-to-end training}
\For{$e \gets 1$ \KwTo $E$}{
  \For{each minibatch $\subset \mathcal{D}^{\mathrm{train}}$}{

    \tcp{Phase 1: Initialization}
    Initialize node and edge hidden states using~\eqref{eq:initializationeq}
    \label{step:init}
    
    \tcp{Phase 2: Message passing}
    \For{$t \gets 0$ \KwTo $T-1$}{
      Compute BS-side attention coefficients $\alpha_{b,u\leftarrow v}^{(t)}$ using~\eqref{eq:bs_attention}, and UE-side attention coefficients $\beta_{b\leftarrow w,u}^{(t)}$ using~\eqref{eq:ue_attention}\;
      Compute BS-side messages $\mathbf{m}_{b,u}^{\mathrm{BS}}$ using~\eqref{eq:bs_message}, and UE-side messages $\mathbf{m}_{b,u}^{\mathrm{UE}}$ using~\eqref{eq:ue_message}\;
      Fuse the two messages according to~\eqref{eq:edge_reduce}\;
      Update the edge hidden states according to~\eqref{eq:edge_update}\;
    }

    \tcp{Phase 3: Readout}
    Construct the contextual features $\mathbf{c}_{b,u}$ according to Section~\ref{subsec:gnn_architecture}\;
    Generate beamforming vectors
    $\mathbf{f}_{b,u}=\varphi^{\mathrm{out}}\!\left(\mathbf{c}_{b,u}\right)$

    Form the digital beamformer at each BS as $\mathbf{F}_b=\big[\mathbf{f}_{b,1},\dots,\mathbf{f}_{b,U}\big]$
    
    Normalize each beamformer to satisfy the per-BS power constraint using~\eqref{eq:normdig}

    \tcp{Loss computation}
    Compute the user rates $\{R_u\}_{u=1}^{U}$ with the predicted beamformers $\{\mathbf{F}_b\}_{b=1}^{B}$ and the corresponding mmWave CSI $\mathbf{H}^{\mathrm{mW}}$\;
    Compute the loss $\mathcal{L}=-R^{\mathrm{sum}}$\;

    \tcp{Parameter update}
    Compute gradients $\nabla_{\boldsymbol{\Theta}}\mathcal{L}$ by backpropagation\;
    Update parameters using the chosen optimizer and learning rate:
    $\boldsymbol{\Theta}\leftarrow \mathtt{Opt}\!\left(\boldsymbol{\Theta},\,\nabla_{\boldsymbol{\Theta}}\mathcal{L},\,\eta\right)$
    
  }
}
\end{algorithm}

\begin{figure*}[!htb]
    \centering
    \small
    \begin{minipage}{0.32\textwidth}
        \centering
        \includegraphics[width=\linewidth]{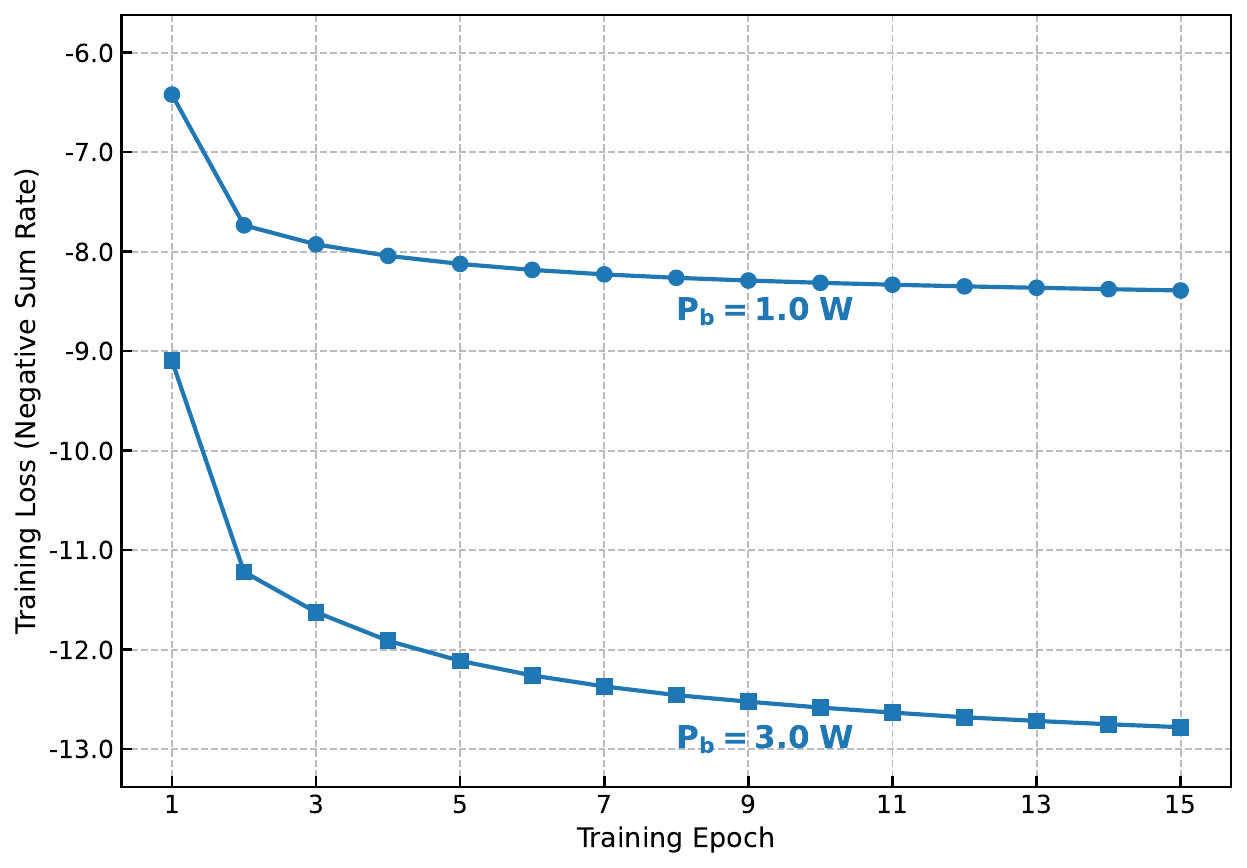}
        \vspace{-3mm}
        \caption{Training loss convergence of the proposed framework for two representative per-BS transmit powers, $P_b\in\{1,3\}$ W.}
        \label{fig:training_loss}
    \end{minipage}\hfill
    \begin{minipage}{0.32\textwidth}
        \centering
        \includegraphics[width=\linewidth]{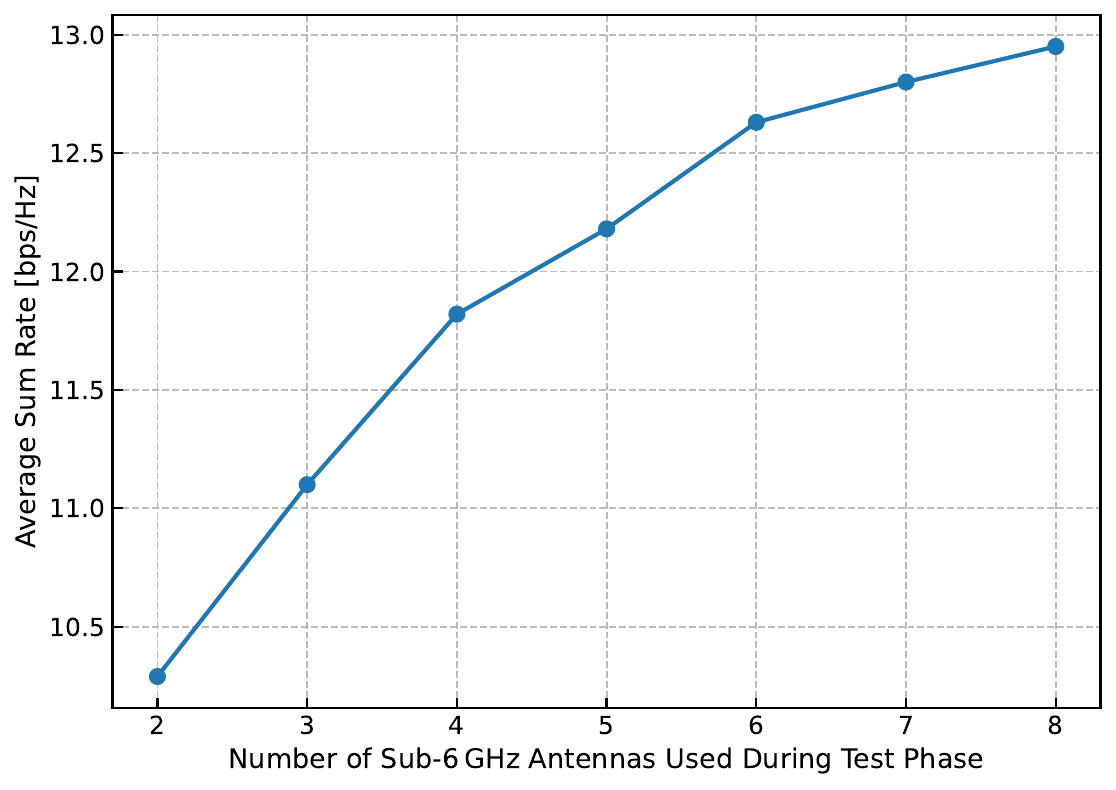}
        \vspace{-3mm}
        \caption{Sum-rate of the GNN-based framework versus the number of available sub-6\,GHz antennas at inference for $P_b=3$ W.}
        \label{fig:SumRate_vs_Sub6GHzAntennas}
    \end{minipage}\hfill
    \begin{minipage}{0.32\textwidth}
        \centering
        \includegraphics[width=\linewidth]{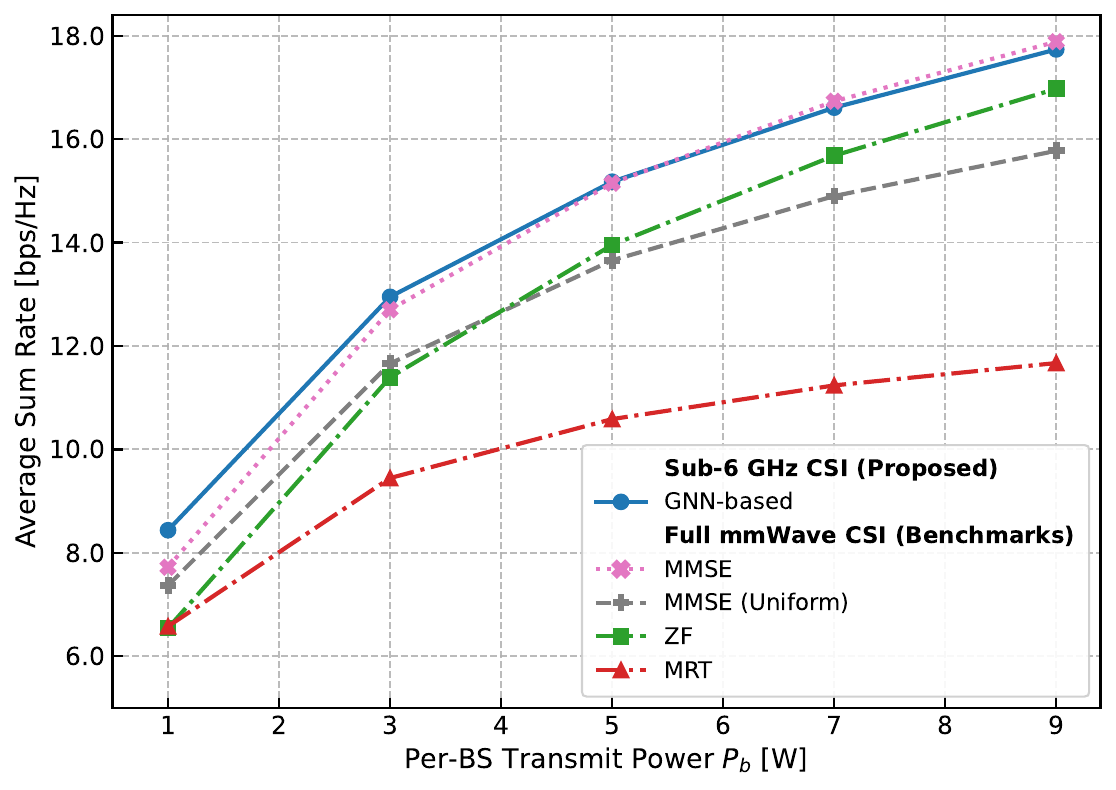}
        \vspace{-3mm}
        \caption{Sum-rate versus per-BS transmit power $P_b$ for the proposed GNN-based beamformer and benchmark beamformers.}
        \label{fig:sumrate_vs_power}
    \end{minipage}
\vspace{-3mm}
\end{figure*}

\section{Numerical Results}
\label{sec_numres}
\textit{Simulation Setup:}
We generate the sub-6\,GHz and mmWave channels using the DeepMIMO dataset~\cite{Alkhateeb2019} under the \texttt{O1} scenario, which is based on ray-tracing simulations obtained from Remcom Wireless InSite. Specifically, we use the \texttt{O1\_28} and \texttt{O1\_3p5} configurations corresponding to the 28\,GHz mmWave band and the 3.5\,GHz sub-6\,GHz band, respectively. We consider $B=3$ BSs that jointly serve the UEs. Each BS employs $N^{\mathrm{mW}}=32$ antennas for the mmWave array and $N^{\mathrm{s}}=8$ antennas for the sub-6\,GHz array, both with half-wavelength spacing. The bandwidths are set to 0.1\,GHz and 0.01\,GHz for the mmWave and sub-6\,GHz bands, respectively, and the number of propagation paths is fixed to $L=15$. 
Using the available sub-6\,GHz CSI, we construct the graphs described in Section~\ref{subsec:graph_construction}. For each graph, we randomly select the number of UEs from the range $[3,8]$ and connect them to all BSs, forming a complete BS-UE bipartite structure. We generate a dataset of $10{,}000$ graphs and randomly split it into training (80\%) and testing (20\%) sets.

We implement the proposed GNN with $T=3$ message-passing layers and hidden dimension $d=64$. 
The attention blocks use the LeakyReLU activation with slope $0.2$. 
The readout module uses a 3-layer MLP with rectified linear unit activations to generate the beamforming vectors.
The model is trained using the Adam optimizer with learning rate of $10^{-4}$ and decay factor of $10^{-5}$. 
We use a batch size of $M=10$ graphs and train the model for $15$ epochs. 

\textit{Convergence Behavior:}
We evaluate the convergence of the proposed GNN-based beamforming training described in Algorithm~\ref{alg:training} in Fig.~\ref{fig:training_loss}. 
The figure shows the training loss over epochs for $P_b=1$ W and $P_b=3$ W. 
In both cases, the training loss decreases rapidly during the first few epochs and gradually stabilizes afterward, indicating stable convergence. 

\textit{Impact of the Number of Sub-6\,GHz Antennas:}
Fig.~\ref{fig:SumRate_vs_Sub6GHzAntennas} shows the downlink sum-rate for $P_b=3$~W versus the number of sub-6\,GHz antennas for which CSI is available during inference.
It is seen that the performance improves consistently as the number of available antennas increases.
For instance, increasing the number of antennas from $3$ to $8$ improves the sum-rate from $11.10$~bps/Hz to $12.95$~bps/Hz. Using only $3$ antennas already achieves about $85.7\%$ of the performance obtained with all sub-6\,GHz antennas.

\textit{Sum-Rate Maximization Results:}
We compare the sum-rate achieved by the proposed GNN-based beamforming scheme with three baselines using full mmWave CSI: zero-forcing (ZF), MMSE, and maximum-ratio transmission (MRT)~\cite{Heath2017MIMOBook}. ZF and MMSE are implemented centrally at the CPU, while MRT is applied locally at each~BS.

Fig.~\ref{fig:sumrate_vs_power} shows the average downlink sum-rate versus the per-BS transmit power $P_b$. All methods use the same GNN-induced per-UE power allocation, where the power assigned by BS~$b$ to UE~$u$ is given by $p_{b,u}^{\mathrm{GNN}}=\|\mathbf f_{b,u}\|_2^2$. The proposed beamformer consistently outperforms ZF and MRT, while remaining close to MMSE.
For example, at $P_b=5$ W, the proposed method achieves $15.18$ bps/Hz, compared to $15.15$ bps/Hz achieved by MMSE, $13.95$ bps/Hz by ZF, and $10.59$ bps/Hz by MRT.
To isolate the benefit of the learned power allocation, Fig.~\ref{fig:sumrate_vs_power} also includes the MMSE beamformer with uniform power allocation, denoted by ``MMSE (Uniform).'' At $P_b=5$ W, the GNN-induced allocation increases its sum-rate from $13.66$ to $15.15$ bps/Hz, corresponding to a $10.9\%$ performance improvement.


\section{Conclusion}
This paper studied fully digital downlink beamforming in CFmMIMO systems using only uplink sub-6\,GHz CSI. We proposed a GNN-based framework that represents the system as a graph and learns beamformers through customized message passing over BS-UE links. Numerical results showed that the proposed approach generalizes well across different network topologies, remains robust under partial sub-6\,GHz CSI at inference, and achieves sum-rate performance close to fully digital benchmark schemes that rely on full mmWave CSI. As future work, this framework can be extended to hybrid beamforming architectures.

\section*{Acknowledgment}
This work was supported by the Research Council of Finland through 6G Flagship Program (grant 369116) and projects DIRECTION (grant 354901), DYNAMICS (grant 24305016), and CHIST-ERA PASSIONATE (grant 359817).

\bibliographystyle{ieeetr}
\bibliography{IEEEabrv,Bibliography}

\vspace{12pt}

\end{document}